\documentclass[%
 reprint,
nofootinbib,
 amsmath,amssymb,
 aps,
floatfix,
]{revtex4-2}
\usepackage{dblfloatfix}
\usepackage[section]{placeins}
\usepackage{graphicx}
\usepackage[labelfont=bf,textfont={bf,normalfont}, justification=raggedright]{caption}
\usepackage{subcaption}
\usepackage[none]{hyphenat}
\usepackage{mathtools}
\usepackage{mathrsfs}
\usepackage{amsmath}
\usepackage{amssymb}
\usepackage{braket}
\usepackage{graphicx}
\usepackage{dcolumn}
\usepackage{bm}
\usepackage{comment}
\usepackage{color,soul}
\usepackage{hyperref}
\hypersetup{
    colorlinks=true,
    citecolor=blue,
    filecolor=blue,      
    urlcolor=blue,
}
\usepackage[normalem]{ulem}

\newcommand{\Cr}{^{48}\text{Cr}}
\newcommand{\Ne}{^{20}\text{Ne}}

\newcommand{\etal}{\emph{et al.}}

\newcommand{\SU}[1]{\ensuremath{\mathrm{SU}( #1 )}}

\newcommand{\SpR}[1]{\ensuremath{\mathrm{Sp}( #1,\mathbb{R} )}}
\newcommand{\hw}{\ensuremath{\hbar\Omega}}

\begin{document}


\title{New Insights into Backbending in the Symmetry-adapted 
Framework}

\author{Nicholas D. Heller$^1$, Grigor H. Sargsyan$^2$, Kristina D. Launey$^2$, Calvin W. Johnson$^3$,  Tom\'{a}\v{s} Dytrych $^{2,4}$, Jerry P. Draayer$^2$}

\affiliation{
 $^1$Department of Physics, Harvey Mudd College, Claremont, CA 91711\\
 $^2${Department of Physics \& Astronomy, Louisiana State University, Baton Rouge, LA 70803} \\
 $^3$Department of Physics, San Diego State University, San Diego, CA 92182 \\
 $^4$Nuclear Physics Institute of the Czech Academy
of Sciences, 25068 \v{R}e\v{z}, Czech Republic}

\date{\today}

\begin{abstract}
We provide new insights into backbending phenomenon within the symmetry-adapted framework which naturally describes the intrinsic deformation of atomic nuclei. For $^{20}$Ne, the canonical example of backbending in light nuclei, the \textit{ab initio} symmetry-adapted no-core shell model shows that while the energy spectrum replicates the backbending from experimental energies under the rigid rotor assumption, there is no change in the intrinsic deformation or intrinsic spin of the yrast band around the backbend.
For the traditional example of $^{48}$Cr, computed in the valence shell with empirical interactions, we confirm a high-spin nucleus that is effectively spherical, in agreement with previous models. However, we find that this spherical distribution results, on average, from an almost equal mixing of deformed prolate shapes with deformed oblate and triaxial shapes. Microscopic calculations confirm the importance of spin alignment and configuration mixing, but surprisingly unveil no anomalous increase in moment of inertia. This finding opens the path toward further understanding the rotational behavior and moment of inertia of medium-mass nuclei.

\end{abstract}
\maketitle

\section{\label{sec:intro}Introduction}

Backbending refers to an anomalous increase in nuclear moment of inertia along the yrast band at some critical angular momentum. Based on experimental energy spectra, backbending is found to occur in nuclei ranging from $\Ne$ through $\Cr$ to the actinide region. 
Understanding the backbending phenomenon is essential to resolve the elusive physics of the high angular-momentum rotational behaviour of strongly-deformed nuclei. Possible explanations of this phenomenon  relate to the physics of spin alignment due to Coriolis force pair-breaking \cite{stephens1975, faessler1976juv}, phase transitions between irrotational and rigid rotor flow, and shape coexistence \cite{ring1980}. Theoretically testing these explanations has been restricted to heavier nuclei, for many of which microscopic $A$-body descriptions
are intractable computationally.

In this paper, we provide the first \textit{ab initio} study of the backbending phenomenon for $\Ne$, a canonical example of backbending in light nuclei \cite{speidel1980}, with a focus on translationally invariant moments of inertia and intrinsic deformation and triaxiality.  
We use \textit{ab initio} wavefunctions for $\Ne$ from our earlier study \cite{dytrychldrwrbb20} calculated in the symmetry-adapted no-core shell model (SA-NCSM) \cite{launeymd_arnps21,dytrychldrwrbb20}. The SA-NCSM provides nuclear wavefunctions in terms of SU(3)$\supset$SO(3) basis states without breaking the rotational symmetry. The SU(3) quantum numbers, in turn, directly provide the intrinsic nuclear deformation in the body-fixed frame \cite{launeymd_arnps21}.
We show that the energy spectrum of $\Ne$ replicates the backbending from experimental energies under the rigid-rotor assumption, but interestingly, the \textit{ab initio} results show no change in the intrinsic deformation or intrinsic spin of the yrast band around the backbend in $\Ne$. To further understand this, we examine spin alignment in low-lying states in $\Ne$.

For the last two decades, the heavier nucleus of $\Cr$ has been a key example of backbending, because it tests the predictions of both mean-field and configuration-interaction methods.
Almost all previous studies of $\Cr$ find an yrast band with an intrinsic prolate deformation before the backbend, and after the backbend, find that $\Cr$ transitions towards sphericity and with a lack of an intrinsic state above the backbend (see for example \cite{velazquez-2000, caurier_1995, tanaka, martinez1996, juodagalvis_1999, juodagalvis_2006, roy2010, gao2011, hara-1999,ljungberg2022nuclear}, and Ref. \cite{herrera2017} for a detailed model comparison). However, when Herrera \etal{} \cite{herrera2017} decomposed configuration-interaction shell-model wavefunctions into components specified by the eigenvalues of the SU(3) second-order Casimir invariant operator, they found 
consistently large deformations 
above the backbend rather than sphericity. While the full SU(3) content of the wavefunction provides information about the intrinsic deformation and rotational bands \cite{launeydd16,bahrir00,dytrychsbdv_prca07,johnson15,johnson16,johnson20,dytrychldrwrbb20}, the second-order Casimir invariant alone is not sufficient to fully decompose into SU(3) irreducible representations (irreps) and distinguish between prolate, oblate, and triaxial deformations. 

To resolve this apparent contradiction about the structure of $\Cr$ after the backbend, we again employ the symmetry-adapted framework. To understand the results of Ref. \cite{herrera2017}, we utilize the same model conditions as those used in \cite{herrera2017} for $\Cr$, namely, a core of inactive particles, valence shell model space, and an empirical interaction for the $pf$ shell. With this, we can now compute $\Cr$ wavefunctions in a configuration-interaction framework with complete information about intrinsic deformation. We show that for the yrast band above the backbend
the deformed configurations observed in Ref. \cite{herrera2017} are a part of a remarkably balanced mixing of prolate and oblate/triaxial intrinsic shapes, which leads to a nucleus that appears spherical on average. 
This reconciles the outcomes of Ref. \cite{herrera2017} and previous studies.

Similarly to $\Ne$, our results for $\Cr$ do not show an anomalous change in moment of inertia at the backbend, suggesting that a
rigid shape change is not the sole mechanism for the backbend. 

The outcomes of our study for $\Ne$ and $\Cr$ emphasize the importance of band crossing and spin-alignment in driving backbending, which affects the energy of the states,
but with only very little effect on the nuclear spatial distribution.

\section{\label{sec:theory}Theoretical Framework}

\subsection{\label{sec:theory_MoI} Backbending and the rigid rotor model}

\begin{figure}[h]
    \includegraphics[width=0.48\textwidth]{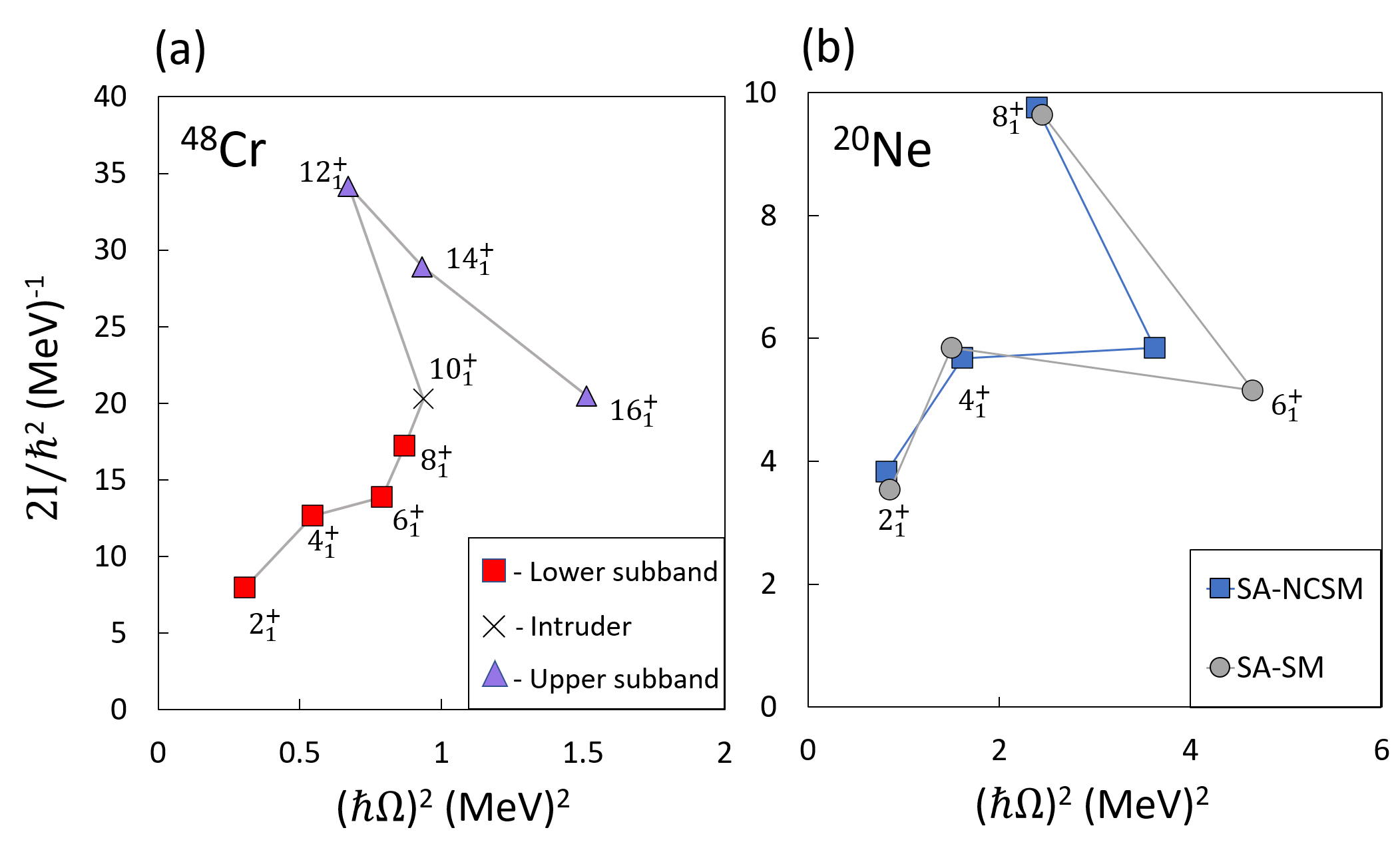}
    \vspace{-.5cm}
    \caption{Traditionally observed backbanding in the yrast band of $\Cr$ at $J=12$ and of $\Ne$ at $J=8$, based on Eq. (\ref{eq:rigid-rotor-MoI}) and excitation energies calculated for: (a) $\Cr$ in the SA-SM with the GXPF1 interaction, and (b) for $\Ne$ in the SA-NCSM with the NNLO$_{\text{opt}}$ chiral potential  for $\hbar \Omega = 15$ MeV and  $N_{\rm max} = 8$ (blue squares), and in the SA-SM with the USDA interaction (gray circles). Excitation energies are in close agreement with experiment, as shown in Fig. \ref{fig:energies}.}
    \label{fig:backbend}
\end{figure}

Backbending is, in principle, observed from applying the rigid rotor model to high-spin energy spectra. The rigid rotor Hamiltonian is
\begin{equation} \label{eq:rr_hamiltonian}
    \hat{H} = \text{const} + \frac{\hbar^2}{2\mathcal{I}} \hat{J}^2,
\end{equation}
with egenvalues $E(J) = E_0 + \frac{\hbar^2}{2\mathcal{I}} J(J+1)$, where $\mathcal{I}$ is the moment of inertia and $\hat{J}$ is the total angular momentum operator. 
Traditionally, the anomalous increase in the moment of inertia has been identified from the dependence of
$2\mathcal{I}/\hbar^2$ on the nuclear rotational frequency $(\hbar \omega)^2$  derived from the rigid rotor excitation energies \cite{stephens1975} (e.g., see Fig. \ref{fig:backbend}): 
\begin{equation} \label{eq:rigid-rotor-MoI}
    \frac{2\mathcal{I}}{\hbar^2} = \frac{4J-2}{E_{\gamma}} \ \ , \ \
    (\hbar \omega)^2 = (J^2 - J + 1)\left(\frac{E_{\gamma}}{2J-1}\right)^2,
\end{equation}
where $E_{\gamma} = E(J) - E(J-2)$ is the excitation energy. In Eq. (\ref{eq:rigid-rotor-MoI}), $\mathcal{I}$ is derived from the first discrete derivative of $E(J)$ with respect to $J(J+1)$, whereas $\hbar \omega$ is derived from the rotational energy at mid-point $\frac{E(J)+E(J-2)}{2} \sim {1 \over 2} \frac{\mathcal{I}}{\hbar^2} (\hbar \omega)^2$ (e.g., see \protect{\cite{el-kameesy2006}}). 

\subsection{\label{sec:theory_SASM} Symmetry-adapted framework}

{\it Ab initio} descriptions of spherical and deformed nuclei up through the calcium region are now possible with the \textit{ab initio} SA-NCSM  without the use of interaction renormalization procedures, as reviewed in Refs. \cite{launeydd16,launeymd_arnps21}. In particular, we have shown that the SA-NCSM, using the SU(3)-adapted basis \cite{dytrychsbdv_prl07} or the \SpR{3}-adapted basis \cite{dytrychldrwrbb20}, can use significantly reduced  model spaces as compared to the corresponding  ultra-large  conventional model spaces without compromising the accuracy of results for various observables. This allows the SA-NCSM to accommodate larger model spaces  and to reach heavier nuclei, such as $^{20}$Ne \cite{dytrychldrwrbb20}, $^{21}$Mg \cite{ruotsalainen19}, $^{22}$Mg \cite{henderson2017dqc}, $^{28}$Mg \cite{physrevc.100.014322},  as well as $^{32}$Ne and $^{48}$Ti \cite{launeymd_arnps21}. 

The SA-NCSM with SU(3)-adapted basis solves the many-body nuclear Hamiltonian in basis states that are labeled schematically as
\begin{equation} 
|\vec{\gamma}; N(\lambda\,\mu)\kappa L; (S_{p}S_{n})S; J M\rangle,  \label{SAbasis} 
\end{equation}
where $S_{p}$, $S_{n}$, and $S$ denote proton, neutron, and total intrinsic
spins, respectively, $N$ is the total number of harmonic oscillator (HO) excitation quanta, and
$(\lambda\,\mu)$ represent a set of quantum numbers that labels an \SU{3} irrep. 
The label $\kappa$ distinguishes
multiple occurrences of the same orbital momentum $L$ in the parent irrep
$(\lambda\,\mu)$.  The $L$ is coupled with $S$ to the total angular momentum
$J$ and its projection $M$.    The symbol $\vec{\gamma}$ schematically denotes the
additional quantum numbers needed to specify a distribution of nucleons over
the major HO shells and their single-shell and  inter-shell quantum numbers.
All of these labels
uniquely determine the SA-NCSM basis states~(\ref{SAbasis}) \cite{launeydd16}.  

For comparatively large model spaces, it is often advantageous to assume a core of inactive particles. For this, we introduce a core in the SA-NCSM and allow only the valence particles to excite to higher shells. This approximate model is henceforth referred to as the symmetry-adapted shell model (SA-SM). When solved only in the valence shell with empirical interactions, it coincides with earlier valence-shell models with SU(3) $\supset$ SO(3) basis \cite{elliott58,verhaar60,harvey68,millener78,hechtZ79,alburgerLOM81,hechtS82,hayes92,beuschelDRH98,vargas01,bahri04su3,kota20}.  For the SA-SM, we transform the effective one-body interactions (Appendix \ref{appSU3}) and two-body matrix elements \cite{launeyddsd15} to the SU(3) basis. 
In this study, we apply the valence-shell SA-SM with empirical interactions to $^{48}$Cr with a $^{40}$Ca core, and to $\Ne$ with a $^{16}$O core. For $\Ne$, we also employ the \textit{ab initio} SA-NCSM for comparisons. 

\subsection{\label{sec:theory_SASM} Moments of inertia and intrinsic deformation}

The SU(3) basis allows calculations of the \textit{microscopic} translationally invariant $\mathcal{I}_z$ moment of inertia for each nuclear state $J$, 
\begin{equation}
 \mathcal{I}_z =  \langle J |\sum_{i=1}^{A} m( r_i^2 - z_i^2 ) | J \rangle, 
\label{microI}   
\end{equation}
where $m$ is the nucleon mass and $A$ the total number of particles.  For each  basis state (\ref{SAbasis}) the quantum numbers $N(\lambda \ \mu)$ 
fully define the HO quanta along the three principal axes: $N_{x}, N_{y}$, and $N_{z}$, with $N=N_{x}+ N_{y}+ N_{z}$, $\lambda = N_z-N_x$, and $\mu=N_x-N_y$. Note that $N$ is given by the total HO energy $N\hw$ of all particles (for a HO frequency \hw), and for example, $N=50$ ($N=156)$ for the valence-shell configurations of $\Ne$ ($\Cr$).
The translationally invariant moment of inertia for a basis state can be then expressed in terms of $N$ and $N_z={1\over 3} (N+2\lambda+\mu)$ (see Appendices \ref{appI} and \ref{appCM}):
\begin{equation} \label{eq:M_o_I}
\frac{2 \mathcal{I}_z^{N(\lambda\, \mu)}}{\hbar^2} = \frac{4}{3{\hbar \Omega}} \left(N - \lambda - \frac{\mu}{2}  - \frac{3}{2}\right).
\end{equation}
We use this equation to calculate the moment of inertia for each SU(3) basis state, and take the probability-weighted sum to find the microscopic moment of inertia of the entire wavefunction. While SU(3) basis states are regarded as rigid rotors, non-rigid degrees of freedom such as spin-coupling or phase transitions manifest themselves through the SU(3) mixing that results from the nuclear interaction.

In addition, the quantum numbers $(\lambda \ \mu)$ of the SU(3) basis states (\ref{SAbasis}) describe the average intrinsic deformation, with $(\lambda \ 0)$, $(0 \ \mu)$, and $(0 \ 0)$ representing pure prolate, oblate, and spherical deformations, respectively.
Following Ref. \cite{bahrir00}, we identify rotational bands as states that exhibit quasidynamical SU(3) symmetry. The corresponding wavefunctions have similar $(\lambda \ \mu)$ decomposition. Alternatively, the wavefunction can be expressed in terms of the $C_2$ eigenvalues of the  second-order SU(3) Casimir invariant, as done in Ref. \cite{herrera2017}, where:
\begin{equation} \label{eq:C2}
C_2(\lambda, \mu) = {\frac{2}{3}}\left(\lambda^2+\mu^2+\lambda \mu + 3\lambda + 3\mu\right).
\end{equation}
Since $C_2$ eigenvalues are symmetric under the exchange of $\lambda$ and $\mu$, 
we will need a second measure to determine if a deformation is prolate ($\lambda > \mu$), triaxial ($\lambda = \mu$), or oblate ($\lambda < \mu$). In this study, we also use the $C_3$ eigenvalues of the third-order SU(3) Casimir invariant operator, proportional to $\lambda - \mu$, which provide exactly this \cite{rosensteel1980}:
\begin{align} \label{eq:C3}
    C_3(\lambda, \mu) = \frac{1}{9}
    (\lambda -\mu)(\lambda +2\mu+3)(2\lambda +\mu+3).
\end{align}

These Casimir invariant eigenvalues can be directly

\onecolumngrid

\begin{figure}[b]
\centering
\includegraphics[width=0.99\textwidth]{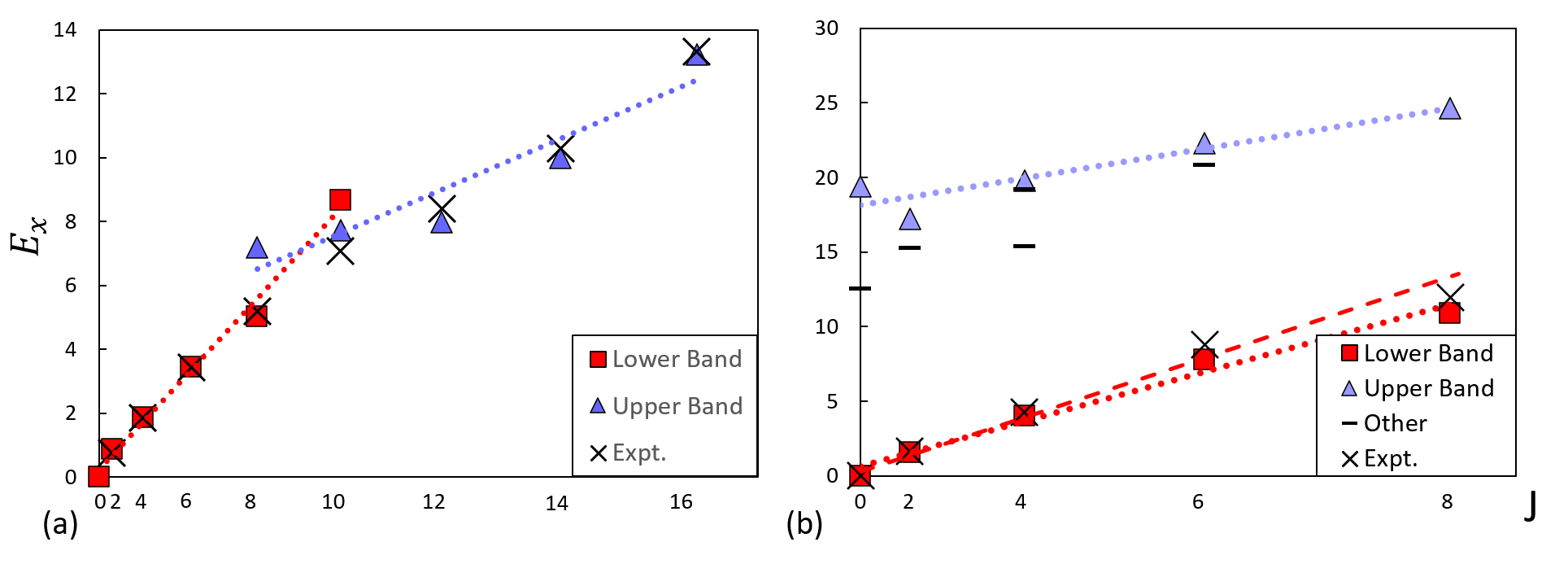}
\caption{Excitation energy $E_x$ vs angular momentum $J$ [scaled by $J(J+1)$ to illuminate rotational bands] for (a) $\Cr$ calculated with the SA-SM with the GXPF1 interaction, and (b) $\Ne$ calculated with the SA-NCSM with the NNLO$_\mathrm{opt}$ chiral potential for $\hbar \Omega = 15$  MeV in $N_{\text{max}} = 8$. The yrast energies are compared to experiment. States are organized in bands according to the dominant SU(3) basis states. Linear regression (dotted line) for each band shows an effective rigid rotor behavior; the linear regression for the lowest four states in $\Ne$ is also shown (dashed line).}
\label{fig:energies}
\end{figure}
\twocolumngrid

\mbox{}
\clearpage
\newpage

\noindent related to the collective shape parameters through the quadrupole moment operator $Q_2=\sum_i \sqrt{16\pi/5} r_i^2Y_2(\hat {\mathbf r}_i)$, where $r_i$ are the particle coordinates relative to the center of mass (CM). Namely, the deformation $\beta$ and  triaxiality $\gamma$ are derived from the expectation values ${1\over 6}\langle Q\cdot Q \rangle  = {3\over 2} k^2 \beta^2$ and $-{1 \over 36} \sqrt{{7 \over 2}}\langle [Q\times Q]_2 \cdot Q \rangle ={3\over4} k^3 \beta^3\cos{3\gamma}$  \cite{mustonengab2018}, where the constant $k=\sqrt{5/9\pi} A r_{\rm rms}^2$, with a root-mean-square (rms) radius $r_{\rm rms}=\sqrt{\langle r^2 \rangle/A}$.
As derived in \cite{castanosdl88}, for a single HO shell, $ k^2 \beta^2 \sim C_2(\lambda,\mu) $ and $ k^3 \beta^3\cos{3\gamma} \sim C_3(\lambda,\mu) $ for an SU(3) basis state, with $ \langle r^2 \rangle = \langle \sum_i r_i^2 \rangle = \frac{\hbar}{m \Omega} (N-3/2)$, where 
$-3/2$ eliminates the CM contribution (see Appendix \ref{appCM}) \cite{tobinflddb14}. For $L^2< 2C_2$, which is often the case, one obtains the relation of $(\lambda\, \mu)$ to the $\beta$ and $\gamma$ shape parameters:
\begin{eqnarray} \label{eq:betagamma}
    &&k \beta \cos \gamma = (2\lambda +\mu +3)/3, \nonumber\\
     &&k \beta \sin \gamma = (\mu +1)/\sqrt{3}.   
\end{eqnarray}
This relation maps each SU(3) basis state with $(\lambda\,\mu)$ to an average ellipsoid with deformation $\beta$ and triaxiality $\gamma$ \cite{castanosdl88,rosensteelr77,leschberD87}. Further mapping the spherical $(0\, 0)$ configuration to $\beta=0$, we obtain $\beta \cos \gamma = (2\lambda +\mu )/3k +\mathcal{O}(1/N)$ and $\beta \sin \gamma = \mu/\sqrt{3}k +\mathcal{O}(1/N)$. According to this relation to the shape parameters we refer to configurations with  $\lambda > \mu$ and $C_3>0$ as prolate deformation ($0^\circ \leq \gamma<30^\circ$) and with $\lambda \leq \mu$ and $C_3 \leq 0$ as oblate deformation ($30^\circ \leq \gamma \leq 60^\circ$), including triaxial ($\gamma=30^\circ$).

\section{\label{sec:results}Results}
We explore the backbending phenomenon by calculating moments of inertia for the yrast band of $^{20}$Ne and $^{48}$Cr, and by examining the symmetry-adapted wavefunctions in terms of their expansion in SU(3) basis states. 

SA-SM calculations for $^{48}$Cr use the $pf$ valence shell, a closed $^{40}$Ca core, and the  GXPF1 empirical interaction \cite{gxpf1} in SU(3) basis. SA-SM calculations for $^{20}$Ne use the $sd$ valence shell, a closed $^{16}$O core, and the  USDA empirical interaction \cite{physrevc.78.064302} in SU(3) basis. 
For $^{20}$Ne, we compare to the \textit{ab initio} SA-NCSM calculations in 11 HO shells ($N_{\text{max}}=8$) of Ref. \cite{dytrychldrwrbb20} with the NNLO$_{\text{opt}}$ chiral potential \cite{ekstrom13} for $\hbar \Omega = 15$ MeV (see Figs. 1 and 3 of Ref. \cite{dytrychldrwrbb20} for wavefunctions and energies, respectively). Using these SA-NCSM calculations, we present the first investigation of backbending and moments of inertia for $^{20}$Ne within an \textit{ab initio} framework.

\subsection{\label{sec:results_energies} Moments of Inertia: Microscopic vs. Energy-spectrum Informed}

We start with the traditional approach to calculating the moment of inertia as a function of the nuclear rotational frequency $(\hbar \omega)^2$ by using the rigid rotor Eq. (\ref{eq:rigid-rotor-MoI}) deduced from excitation energies. Indeed, the SA-SM and SA-NCSM calculations yield energies that reproduce the well-known backbends of $\Cr$ and $\Ne$ (Fig. \ref{fig:backbend}). 

Likewise, according to Eq. (\ref{eq:rr_hamiltonian}), moments of inertia can be extracted within a rotational band if the energy of its states follow the rigid rotor $J(J+1)$-dependence, as shown in Fig. \ref{fig:energies} (cf. Ref. \cite{herrera2017} for $\Cr$). In the SA framework, we organize states into rotational bands according to the dominant SU(3) basis states \cite{launeydd16}. To guide the eye, a linear regression for each of the rotational bands identified in Fig. \ref{fig:energies} provides a slope that is
\onecolumngrid

\begin{figure}[h]
    \includegraphics[width=0.8\textwidth]{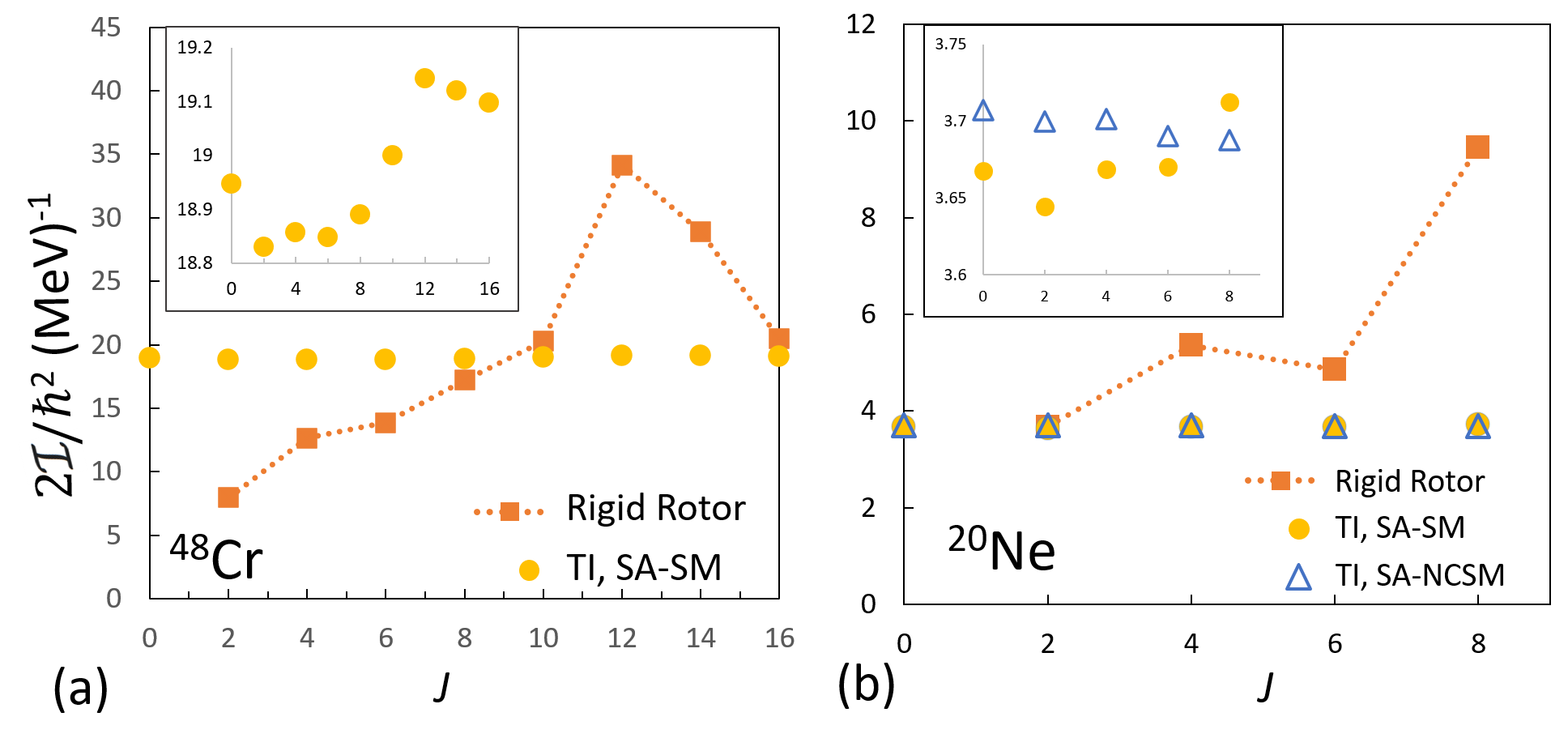}
    \caption{Moment of inertia $2 \mathcal{I}_z / \hbar^2 $ vs angular momentum $J$ for the yrast bands of (a) $\Cr$ and (b) $\Ne$. Translationally invariant (TI) microscopic moments of inertia use Eq. (\ref{eq:M_o_I}) and are calculated in the SA-(NC)SM framework (``TI, SA-SM" and ``TI, SA-NCSM"), whereas the deduced moments of inertia (``Rigid Rotor") use the rigid rotor Eq. (\ref{eq:rigid-rotor-MoI}) and experimental energies. Insets: the same without the rigid rotor deduced values, showing the slight variations in moments of inertia.}
     \label{fig:microMoI}
\end{figure}
\twocolumngrid
\noindent
inversely proportional to the average moment of inertia. In addition, we report the experimental energies of the yrast band, which are in close agreement with the calculations. 

Specifically, calculated excitation energies for $^{48}$Cr replicate the well-known crossing of two rotational bands: the lower band that starts at the ground state, and the upper  band that exhibits strong rigid rotor behavior with a shallower slope, suggesting doubling of the moment of inertia
 (Fig. \ref{fig:energies}a). 
The upper band first appears as an excited state at $J=8$, and becomes the yrast band  at $J=10$. This crossing marks the backbend of $^{48}$Cr as seen in the dramatic shift in moment of inertia (Fig. \ref{fig:backbend}a).
Similarly for $^{20}$Ne (Fig. \ref{fig:energies}b), the yrast band displays strong rigid rotor behavior from $J=0$ to $J=6$. We note that in Fig. \ref{fig:energies}b, the linear regression of the first four states only (dashed line) has a steeper slope than the one up to $J=8$ (dotted line). Hence, the energy of the lowest $8^+$ state is lower compared to the rigid rotor prediction, producing an apparent backbend.
\begin{figure}[th]
    \includegraphics[width=1\columnwidth]{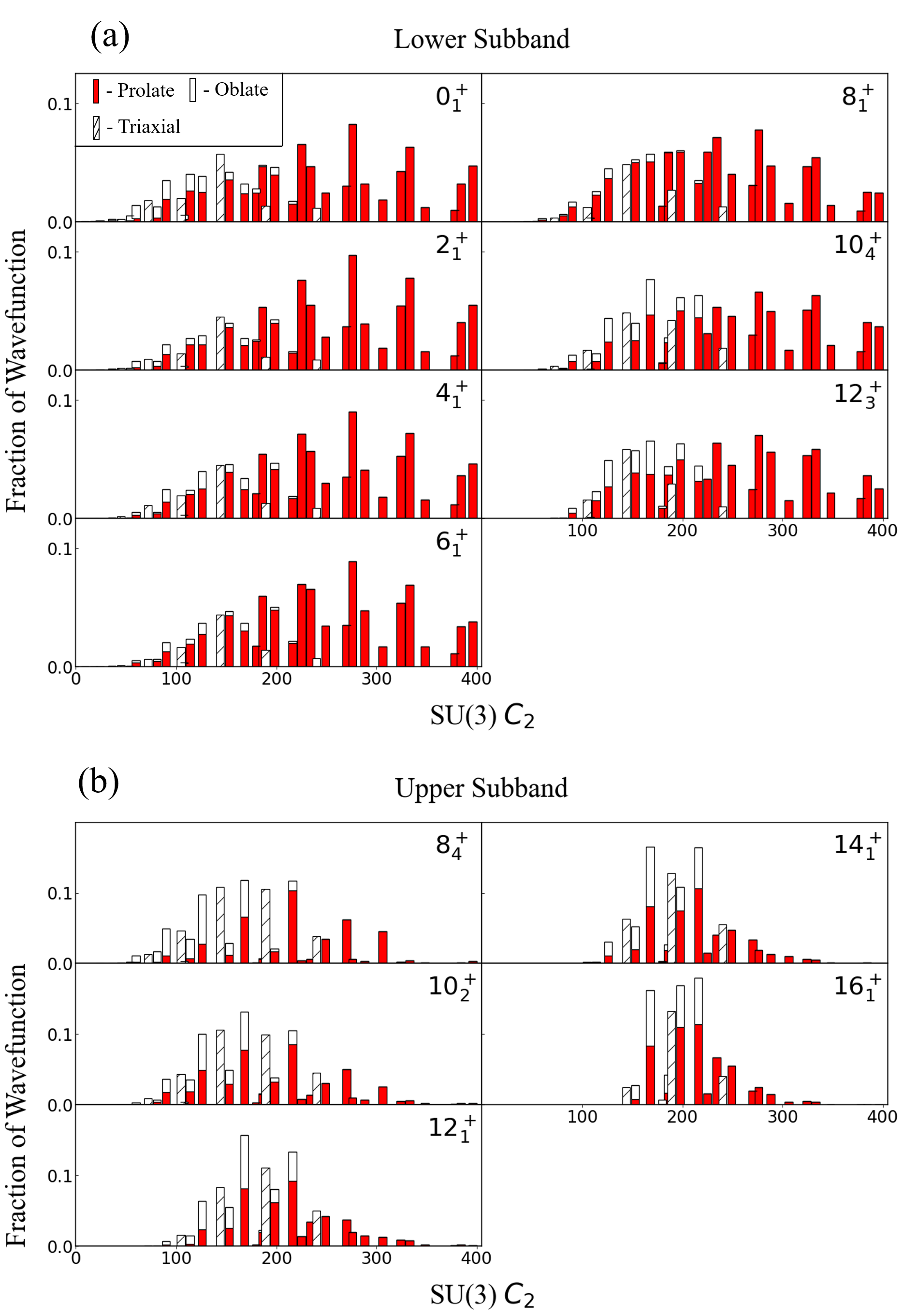}\\
    \raggedright \hspace{0.28in} (c) \\
    \includegraphics[width=0.93\columnwidth]{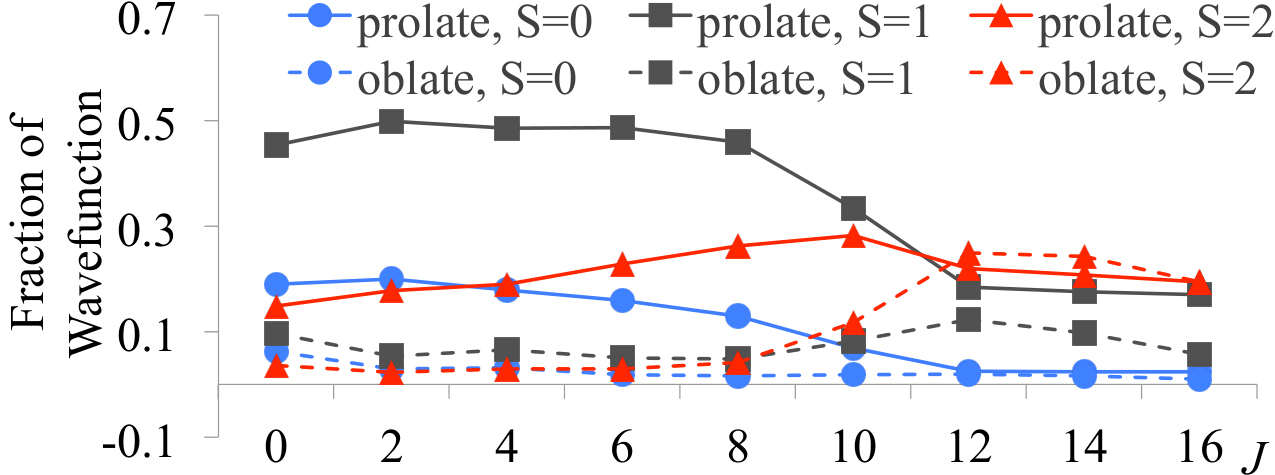}
    \caption{Decomposition of states in $\Cr$ with GXPF1 across deformation $\beta$ ($x$-axis), or equivalently $\mathcal{C}_2={3 \over 2}C_2$ (with the most deformed configurations having the largest $C_2$) and triaxiality $\gamma$ (stacked bars: filled red for prolate, unfilled for oblate, and diagonally-striped  for triaxial deformation). (a) Lower subband. (b) Upper subband starting at $J=8$. (c) Spin decomposition of the yrast states (circles for spin zero, squares for spin one, and triangles for spin two) for prolate (solid) and oblate (dashed) deformations. }
    \label{fig:Cr_wavefunctions}
\end{figure}

Remarkably, microscopic calculations of the moment of inertia show no anomalous increase for both nuclei (Fig. \ref{fig:microMoI}). We calculate microscopic moments of inertia using Eq. (\ref{microI}) by taking into account the position of each particle and with a proper treatment of the center-of-mass. Microscopic moments of inertia of the yrast band have a magnitude similar to the ones deduced from the experimental energy spectra near and below the backbend, but remain practically unchanged after the backbend. This is also confirmed by the \textit{ab initio} wavefunctions in $\Ne$  ($\mathcal{I}_z \approx 3.7 \ \text{MeV}^{-1}$) that do not predict large moment of inertia  for $8^+_1$ in contrast to the rigid rotor results. We note that the SA-SM results suggest an increase in $\mathcal{I}_z$, which however is only  1.3\% for the upper band compared to the lower band in $\Cr$ and only 1.2\% for $J=8$ in $\Ne$ (see the inset of Fig. \ref{fig:microMoI}). This increase is likely the effect of the very small number of configurations available for high $J$ in the valence shell, thereby reducing mixing.
To explain these findings, we next examine the intrinsic structure of the yrast states in $\Ne$ and $\Cr$.

\subsection{\label{sec:results_wavefunctions} Intrinsic Deformation and Spin of Nuclear States}

The symmetry-adapted basis naturally provides decomposition into $(\lambda \ \mu)$ configurations that inform about the intrinsic deformation. From $(\lambda \ \mu)$ we can calculate $C_2$ eigenvalues (\ref{eq:C2}) as well as $C_3$ (\ref{eq:C3}), along with the shape parameters, $\beta$ and $\gamma$ (\ref{eq:betagamma}).  For $\Cr$, this expands the $C_2$ analysis of Ref. \cite{herrera2017} and allows us to gain insight into the type of deformation, namely, prolate, oblate or triaxial. Furthermore, we discuss features important to understand the backbending phenomenon by examining $\Ne$ \textit{ab initio} wavefunctions.  

\begin{figure}[t]
 \includegraphics[width=0.49\textwidth]{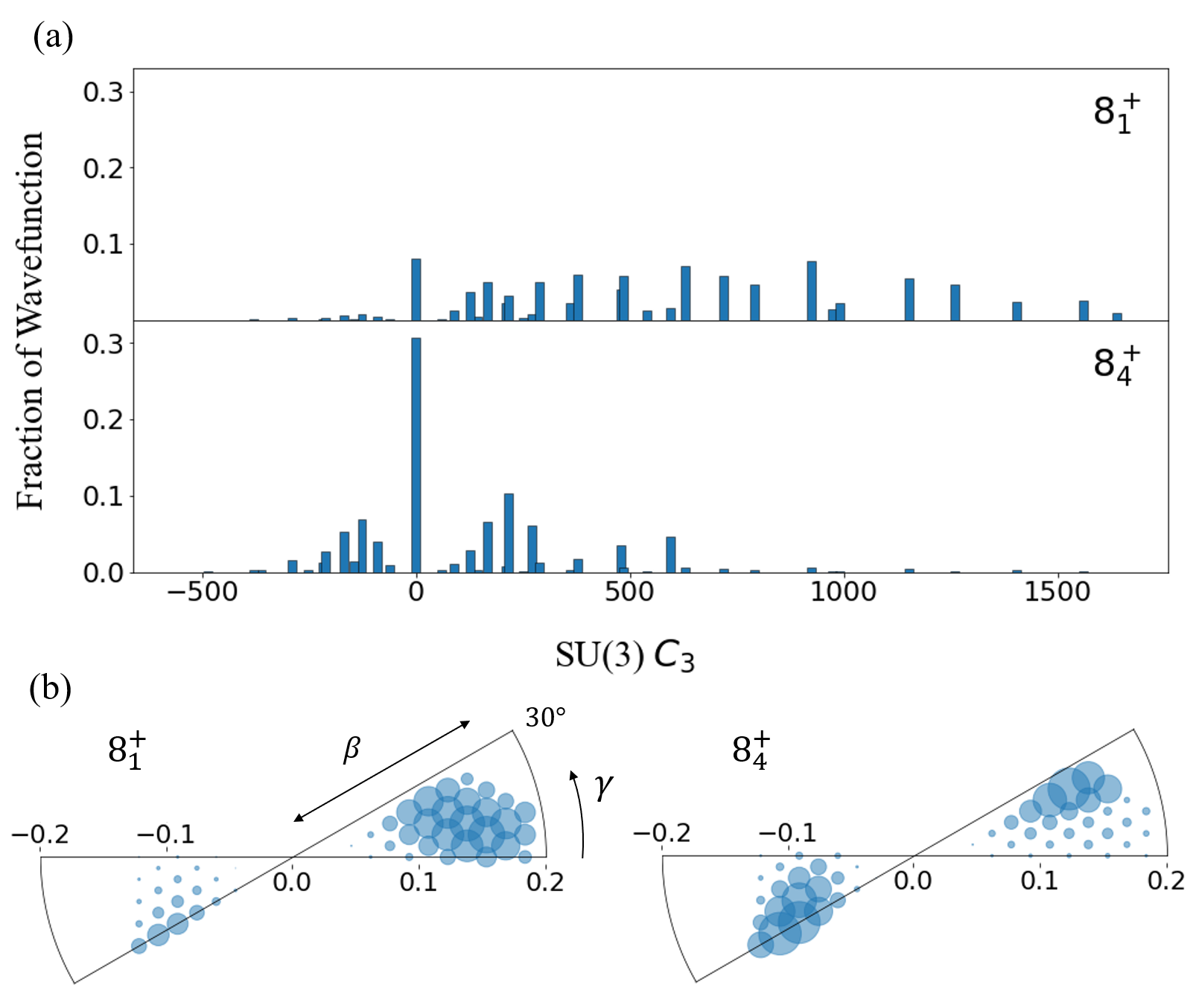}
 \caption{Comparison of the lower and upper subbands of $\Cr$ at $J=8$. Wavefunction decomposition across (a) SU(3) $C_3$ (with $C_3>0$ for prolate, $C_3=0$ for triaxial, and $C_3<0$ for oblate deformation), and (b) $\beta$ and $\gamma$ (no effective charges are used). In (b), the area of each circle is proportional to probability and the origin corresponds to the spherical $(\lambda \,\mu)=(0 \,0)$.}
 \label{fig:Cr_lower_upper_subband_c3}
\end{figure}

\subsubsection{Intrinsic structure of $\Cr$}

Based on the SU(3) content, we group states in $\Cr$ into two rotational bands. Confirming the results of Ref. \cite{herrera2017}, the  ground state band appears as the yrast band of $\Cr$ from $J=0$ to $J=8$ and continues through $J=12$ (Fig. \ref{fig:Cr_wavefunctions}a). Here, we show that the GXPF1 renders this band as strongly prolate (see the filled red bars at large $C_2$ values in Fig. \ref{fig:Cr_wavefunctions}a). We recall that large $C_2$ values correspond to large deformation $\sqrt C_2 \sim \beta$ according to Eq. (\ref{eq:C2}).
At $J \gtrsim 10$ near the ``backbend'', we find an increased mixing of oblate and triaxial deformations, likely due to the crossing of the upper rotational band (upper subband). 

Although the upper subband still contains large deformations as seen in the $C_2$ decomposition in Fig. \ref{fig:Cr_wavefunctions}, many of these are oblate with $C_2$ values equal to dominant prolate deformations but with opposite $C_3$ values (see Fig. \ref{fig:Cr_lower_upper_subband_c3} for the difference between the lower and upper subbands at $J=8$). Without knowing $C_3$, the upper band would have been associated with a still overall deformed shape, as suggested in Ref. \cite{herrera2017}. The $C_3$ decomposition can explain this unexpected result, namely, that the yrast band makes a rapid transition at the ``backbend'' to a strong mixing of prolate and oblate deformations, which on average appears spherical, but without intrinsic sphericity. 

To quantify these results, we calculate $\beta$ across each band by using the common convention of positive $\beta$ for prolate  ($\lambda > \mu$), and negative $\beta$ for oblate deformation ($\lambda \leq \mu$) (including triaxial). The lower subband is found to have practically the same $\langle \beta \rangle$ for all its states, with an average value of $\langle \beta \rangle = 0.1$ and a very large $\langle C_3 \rangle = 603.16 $. For each of these states, the largest deformation is $\beta \sim 0.2$, which is expected to double in a complete model space \cite{rowe_book16}. Note that, e.g., for $0^+_1$, the  single (16\, 4) configuration with the largest $C_2$ is the most dominant and its contribution is expected to increase  in larger model spaces, while the peak in Fig. \ref{fig:Cr_wavefunctions}a (see $0^+_1$) occurs at comparatively  smaller $C_2$ and corresponds to 65 different (14\, 2) configurations of various spin values. The dominant configurations in the lower subband correspond to largely prolate intrinsic deformation, as also observed in Fig. \ref{fig:Cr_lower_upper_subband_c3}b for $8^+_1$.   Whereas for the upper subband, we find relatively smaller values for $\langle \beta \rangle = -0.003$ and $\langle C_3 \rangle = 145.39$ (see Fig. \ref{fig:Cr_lower_upper_subband_c3}b for $8^+_4$). This suggests that the upper subband possesses deformed intrinsic states of prolate and oblate/triaxial deformation, which on average appear spherical.

The SA basis also provides information about the intrinsic spin. As shown in Ref. \cite{herrera2017}, the $S=2$ contribution doubles after the ``backbend". Our findings show that the contribution of the $S=2$ prolate deformations practically remains the same (Fig. \ref{fig:Cr_wavefunctions}c). Notably, this increase in the $S=2$ contribution is a result of the new oblate/triaxial configurations. Overall, there is an equal mixing of prolate and oblate deformations across different $S$ values in the upper subband.

\subsubsection{Intrinsic structure of $\Ne$}

\begin{figure}[b]
    \centering
    \includegraphics[width=0.48\textwidth]{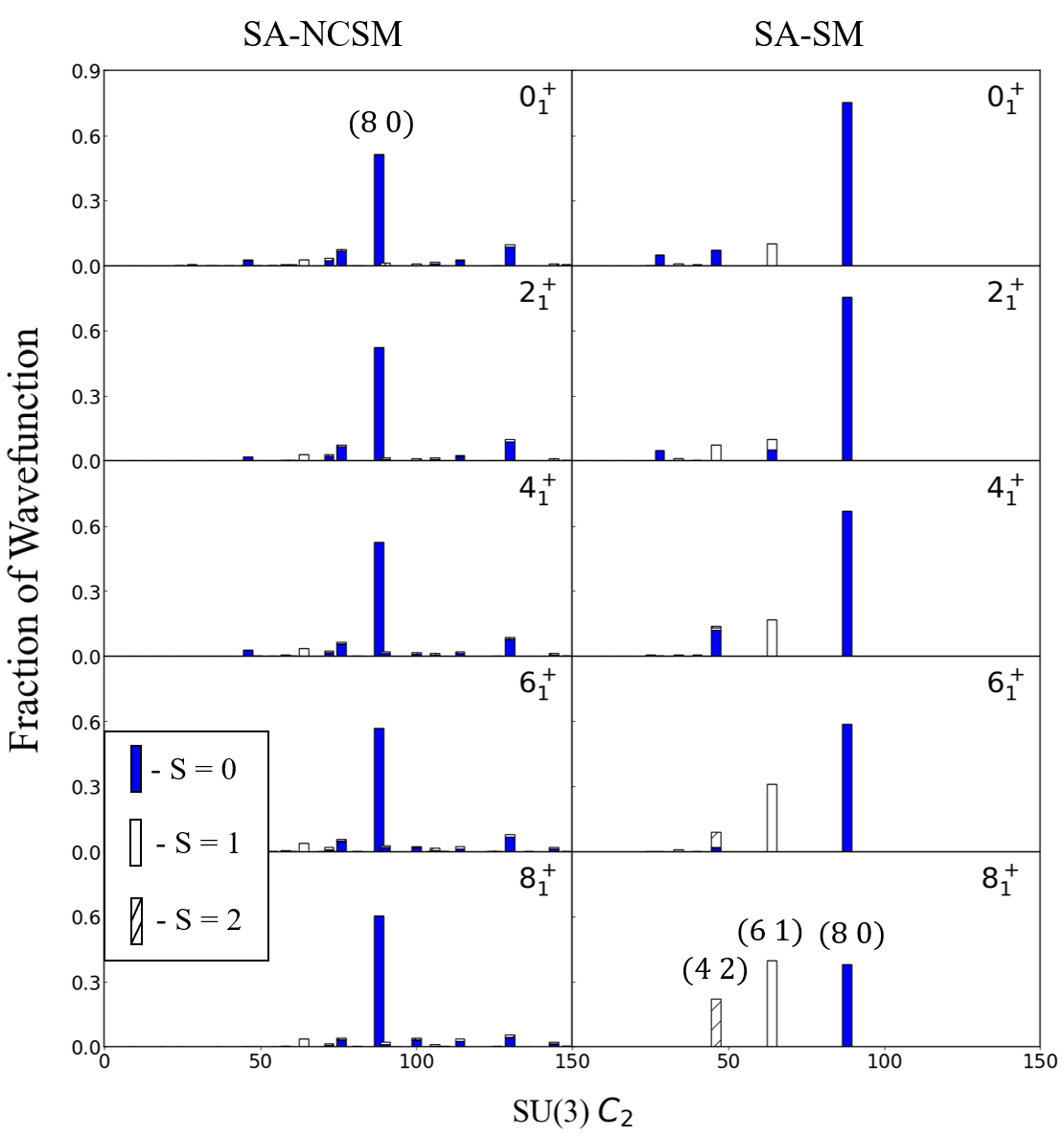}
    \vspace{-.4cm}
    \caption{SU(3) $C_2$ eigenvalue ($\mathcal{C}_2={3 \over 2}C_2$) and spin decomposition for the yrast band states of $\Ne$. (Left) SA-NCSM with the NNLO$_{\text{opt}}$ chiral potential, $N_{\text{max}} = 8$, and $\hbar \Omega = 15$ MeV. (Right) SA-SM with the USDA effective interaction. }
    \label{fig:20Ne_lower_comparison}
\end{figure}

As discussed above, the \textit{ab initio} SA-NCSM results for $\Ne$ showed no increase in moment of inertia from $J=0$ to the ``backbend'' at $J=8$ (Fig. \ref{fig:microMoI}b), which is also supported by the $(\lambda\, \mu)$ and spin decomposition as shown in Fig. \ref{fig:20Ne_lower_comparison}a. Indeed, the lowest five states belong to a single  rotational band that is prolate. The dominant deformation for this rotational band is $(8 \ 0) \ S = 0$, accounting for $50 \%$ to $60 \%$ of the wavefunction, as has been found in previous studies (e.g., see \cite{kazama1988, launeydd16, zbikowski2021}). 

Without a change in nuclear structure and microscopic moment of inertia, what produces the apparent anomaly seen in Fig. \ref{fig:backbend} for $\Ne$? Previous studies of Ne isotopes have pointed to rotational alignment as the particles in the \textit{sd} shell align their angular momenta along the rotational axis at the backbend, resulting in an oblate nucleus rotating around its symmetry axis \cite{szanto,speidel1980,speidel1983}. This mechanism is identical to that commonly used to explain backbending in heavy nuclei \cite{stephens1975}. However, this conflicts with the \textit{ab initio} $\Ne$ wavefunctions that reveal highly prolate deformation across $J=0-8$. With no change in intrinsic structure or angular momentum, the cause of the ``backbend" in $\Ne$ cannot be found looking at the yrast band in isolation as in $\Cr$.

\begin{figure}[t]
    \centering
    \includegraphics[width=0.48\textwidth]{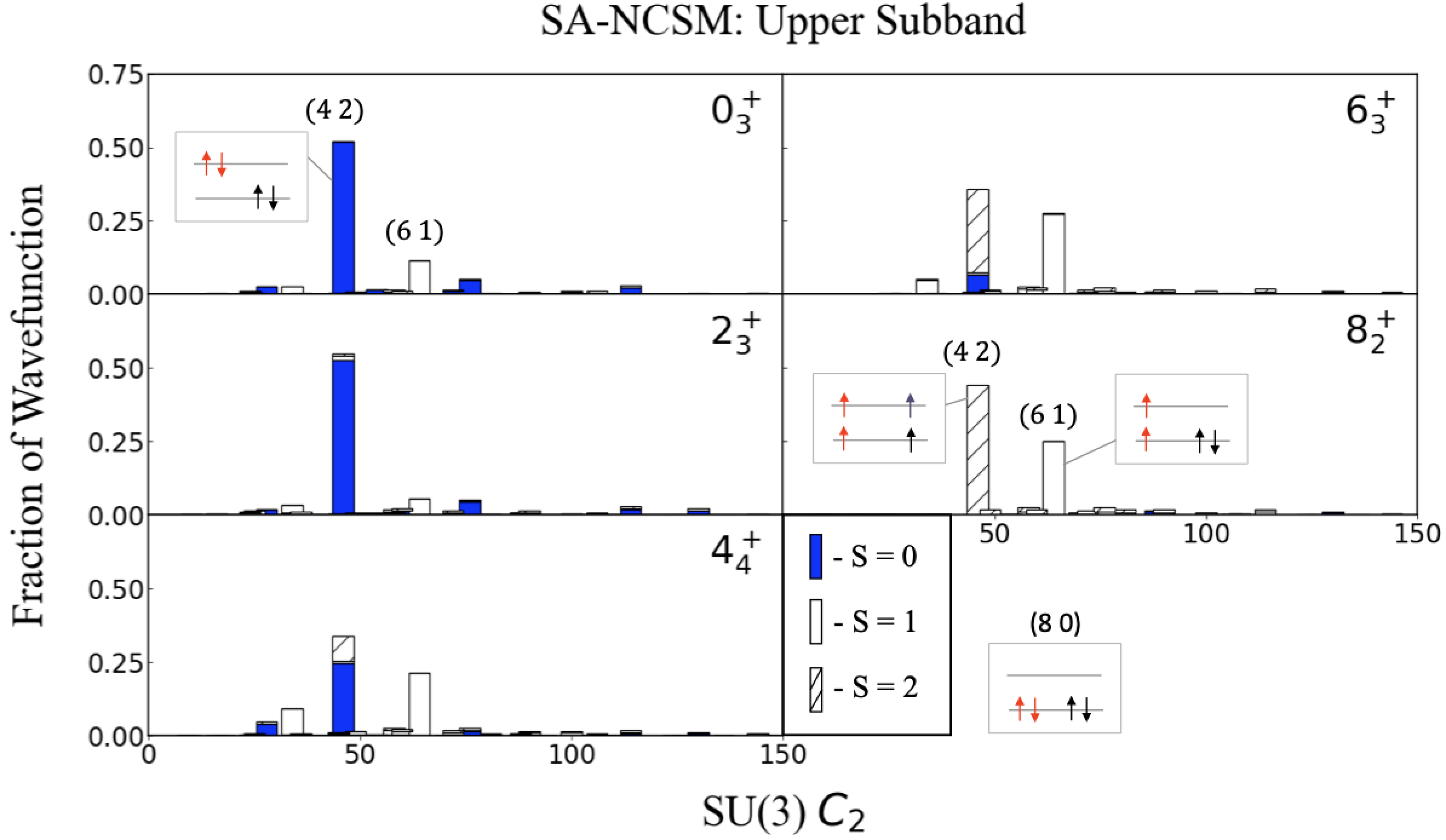}
    \vspace{-.4cm}
    \caption{SU(3) $C_2$ eigenvalue ($\mathcal{C}_2={3 \over 2}C_2$) and spin decomposition of the upper rotational band states of $\Ne$  calculated in the SA-NCSM with the NNLO$_{\text{opt}}$ chiral potential, $N_{\text{max}} = 8$, and $\hbar \Omega = 15$ MeV. Insets:  Spatial distribution of particles within the dominant SU(3) configurations (excluding mirror $p \leftrightarrow n$ configurations). Each level corresponds to specific single-particle excitations in the $sd$ shell and can have maximum of four particles [two protons (red) and two neutrons (blue), with spin up ($\uparrow$) and down ($ \downarrow$)]. For comparison, the most dominant (8\, 0) of the yrast band is also shown.}
    \label{fig:20Ne_upper}
\end{figure}

Here, the SA-SM results provide a hint (Fig. \ref{fig:20Ne_lower_comparison}b). Different from the \textit{ab initio} SA-NCSM, the effective SA-SM does not show a constant $(8 \ 0)$ deformation, but instead displays this deformation diminishing toward the ``backbend'', while the $(6 \ 1) \ S = 1$ and $(4 \ 2) \ S = 2$ configurations increasingly mix. In fact, at $J=8$ the $(6 \ 1) \ S = 1$ deformation slightly dominates over the $(8 \ 0) \ S = 0$. It is interesting to point out that in the SA-SM, there are only three unique intrinsic deformations in the valence shell for $8^+_1$, whereas in the SA-NCSM for 11 shells, there are 116 unique intrinsic deformations. Mixing very similar to the one found with the SA-SM is observed in the no-core shell-model results of Ref. \cite{zbikowski2021} with a different chiral potential, larger $\hw$ and smaller model spaces compared to the results presented in Fig. \ref{fig:20Ne_lower_comparison}a. We note that our SA-NCSM model space includes all the configurations used in Ref. \cite{zbikowski2021} for $\Ne$; in addition, we include selected configurations in  higher shells with no limitations on $(6 \ 1)$, $(4 \ 2) $, and many other deformations compared to $(8 \ 0)$.  Indeed, we find that large model spaces are necessary to develop large deformation (equilibrium shapes and their vibrations), such as the highly prolate $(8\, 0)$ deformation in $\Ne$, and to reduce mixing \cite{launeydd16}. This is further confirmed by the SA-NCSM results in Fig. \ref{fig:20Ne_lower_comparison}a.

The SA-SM large mixing is reminiscent of the lower rotational band of $\Cr$ which increasingly mixes with the dominant shapes of the upper rotational band around the ``backbend". With this resemblance, we can expect another distinct rotational band lying above the yrast band of $\Ne$. 

Indeed, in the SA-NCSM framework, we find that the the second excited state of $J=8$ ($8_2^+$) is dominated by the two deformations $(4 \ 2) \ S = 2$ and $(6 \ 1) \ S = 1$ (Fig. \ref{fig:20Ne_upper}), the same deformations we see increasingly mixing in the SA-SM results of Fig. \ref{fig:20Ne_lower_comparison}. We identify this $8_2^+$ state as belonging to a rotational band that extends to $J=0$ as shown in Fig. \ref{fig:energies}b. 

From this upper subband, we make two primary observations. First, the excited states of this band are either the third or fourth excited state at each value of $J$. That is, until the ``backbend'' at $J=8$, where the $8_2^+$ lies directly above the yrast band. Second, approaching the ``backbend'', we find that $(6 \ 1) \ S=1$ and $(4 \ 2) \ S=2$ configurations increasingly dominate. $(6 \ 1) \ S=1$ is also the most dominant $S=1$ configuration in the lower subband (Fig. \ref{fig:20Ne_lower_comparison}a). For the dominant intrinsic deformation $(4 \ 2)$, the $S = 2$ configuration starts to dominate over the $S = 0$ configuration at $J=6$. 

These two dominant deformations are the lowest spatial excitations, with $(4 \ 2) \ S = 2$ representing a proton-proton and neutron-neutron spin-alignment in the $sd$ shell, and with $(6 \ 1) \ S = 1$ representing a proton-proton (or neutron-neutron) spin-alignment (Fig. \ref{fig:20Ne_upper}, insets\footnote{
The single-particle HO basis can be specified by $\ket{n_z n_x n_y}$, the HO quanta in the three Cartesian directions, $z$, $x$, and $y$, with $n_x + n_y + n_z=n$ ($n=0, 1, 2, \dots$ for $s$, $p$, $sd$, ... shells). 
For a given  HO major shell,  the complete shell-model space is then specified by all distinguishable distributions of $n_z, n_x$, and  $n_y$ (see, e.g., \cite{launeydd16}). E.g., for $n=2$, there are 6 different distributions, $(n_z,n_x,n_y)=(2,0,0),(1,1,0),(1,0,1),(0,2,0),(0,1,1)$ and $(0,0,2)$ (the first two configurations are depicted as levels in the insets of Fig. \ref{fig:20Ne_upper}). 
}). 
In other words, we find  spin-alignment in the upper rotational band just as it begins to interfere with the yrast band. We also note that $(4\, 2)$ remains the most dominant configuration throughout the upper subband, thereby yielding practically unchanged moment of inertia (spatial degrees of freedom), but its spin structure changes from $S=0$ at low $J$ to $S=2$ for high $J$. 

\vspace{8pt}
The outcome of this study suggests that for $\Ne$, as well as for the heavier $\Cr$, spin-aligned configurations at high $J$ interfere with  more deformed $S=0$ configurations. This affects the energy of the states, but with only very little effect on the total moment of inertia and nuclear spatial distribution, in contrast to what the energy spectra in Fig. \ref{fig:energies} suggest.  This further corroborates the outcome of Ref. \cite{rowe_prc2020}, where the details of a potential in addition to the rotational energy of a rigid rotor have been found important to reproduce the experimental energies  of the $^{166}$Er ground state rotational band.

\section{\label{sec:conclusions}Conclusions}

While energy spectra from both the SA-SM and SA-NCSM replicate the backbending from experimental energies under the rigid-rotor assumption, our microscopic calculations of moment of inertia do not predict a dramatic change in moment of inertia at the backbends of $\Ne$ and $\Cr$. Instead, these results suggest that band-crossing and spin-alignment may significantly affect the energies but only marginally affect intrinsic deformation along the yrast band.

For $\Cr$ with the GXPF1 effective interaction, we reconcile contradicting predictions of the intrinsic structure after the backbend. Namely, an almost equal mixing of prolate and oblate/triaxial intrinsic deformations leads to a nucleus appearing spherical on average.
However, the overall change in deformation from a strongly prolate low-$J$ structure to this mixture of deformed states leads to only a 1.3\% increase in the microscopic moment of inertia.

In $\Ne$, the traditional rigid rotor ``backbend" is reproduced, however, without any change in intrinsic structure along the yrast band and in the microscopic moment of inertia. We instead find evidence that a spin-aligned upper rotational band mixing with the ground state band leads to a divergence from rigid rotor behavior and has an effect on the excitation energy not on the spatial distribution. 

These outcomes do not support
rigid shape change as a sole mechanism for the backbend of $\Cr$ as in \cite{tanaka,szanto,caurier_1995}, while emphasizing the role of band crossing and spin-alignment.

Our comparison of the SA-SM and SA-NCSM results in $\Ne$ points to the need to 
further explore $\Cr$ in larger model spaces that are computationally intensive.  Primarily, they suggest that the large mixing of deformed states we see in $\Cr$ with GXPF1 may decrease if the condition of the valence shell is relaxed and excitations to higher shells are included, such as in the SA-NCSM. This is clearly seen in the $8^+$ yrast state in $\Ne$, where the valence-shell calculations result in a strong mixing in deformation and spin. In contrast, the larger model space in the SA-NCSM allows for the most deformed shape to develop and become dominant. Since \textit{ab initio} SA-NCSM calculations are feasible in the region of $\Cr$, future work to study $\Cr$ from first principles will provide further insight.

\begin{acknowledgments}
This work was supported in part by the U.S. National Science Foundation  (PHY-1913728), the Czech Science Foundation (22-14497S), and the U.S. Department of Energy (SC0019521). We thank the National Science Foundation for supporting this work through the REU Site in Physics \& Astronomy (NSF Grant \#1852356) at Louisiana State University. This work benefited from high performance computational resources provided by LSU (www.hpc.lsu.edu),  the National Energy Research Scientific Computing Center (NERSC), a U.S. Department of Energy Office of Science User Facility operated under Contract No. DE-AC02-05CH11231, as well as the Frontera computing project at the Texas Advanced Computing Center,  made possible by National Science Foundation award OAC-1818253. This material is based upon work supported by the U.S.
Department of Energy, Office of Science, Office of Nuclear
Physics, under Grant No. DE-FG02-03ER41272.
\end{acknowledgments}

\appendix{
\section{Transformation to SU(3)}\label{appSU3}
For completeness, we present the interaction matrix elements used in the SA-SM. In order to introduce a core in the SA-NCSM, we transform the single-particle energies to the SU(3) basis. After expanding the elements in terms of Clebsch-Gordan coefficients, we reduce them to a sum over the orbital and total angular momentum quantum numbers, $\ell$ and $j$. For example, one-body SU(3) matrix elements are calculated as
\begin{equation}
\begin{aligned}
\varepsilon_{n(\lambda \mu)\kappa (L S) J =0 M =0} &= (-1)^{L} \sqrt{2L+1} \sum_{\ell j} (-1)^{1/2+j} \varepsilon_{n \ell j} \\ & \hspace{-0.5cm} \times (2j+1) C^{(\lambda \mu) \kappa L}_{(n 0)\ell,(0 n) \ell}
\times
\begin{Bmatrix}
   1/2 & \ell & j \\
   \ell & 1/2 & L
\end{Bmatrix}
\end{aligned},
\end{equation}
where $\varepsilon_{nlj}$ are single-particle energies and $C^{(\lambda \mu) \kappa L}_{(n 0)\ell,(0 n) \ell}$ are SU(3) reduced Clebsch-Gordan coefficients (with outer multiplicity $\rho=1$). A similar transformation yields the two-body matrix elements \cite{launeyddsd15}.

\section{Microscopic Moments of Inertia}\label{appI}
The translationally invariant $z$-component of the moment of inertia $\mathcal{I}_z$ is calculated for a given nuclear state $\ket{J^\pi}$ as:
\begin{equation}
    \mathcal{I}_z = m \big\langle J| \sum_{i=1}^A (  x_i^2 +  y_i^2 ) |J \big\rangle = m \langle \sum_{i=1}^A (  r_i^2 -  z_i^2 ) \rangle, \label{momz}
\end{equation}
where ${\mathbf r}_i$ is the coordinate of the $i$th particle relative to the center-of-mass (for simplicity of notations, we will omit the state parity $\pi$ and use expectation values $\langle ... \rangle$).

The SA-SM and SA-NCSM use laboratory-frame coordinates, and we will first derive $\mathcal{I}_z^{\rm L}$ for the laboratory frame (L). We will show the steps to remove the center-of-mass contribution in the next section. 

In terms of the symplectic $\text{Sp}(3, \mathbb{R})$ generators \cite{launeydd16} and the oscillator length $b = \sqrt{\frac{\hbar}{m \Omega}}$, the operators needed to calculate $\mathcal{I}_z^{\rm L}$ are give as
\begin{subequations} \label{eq:ri2}
\begin{align}
    \frac{1}{b^2} \sum_{i=1}^A  ({r}_i^{\hspace{0.05cm} \rm L})^2 &= \sqrt{\frac{3}{2}} \left( A^{(20)}_0 + B^{(02)}_0 \right) + H^{(00)} \notag \\
    &= \sum_{\alpha=x,y,z} (A_{\alpha \alpha} + B_{\alpha \alpha} + C_{\alpha \alpha}),
\end{align}
\end{subequations}
where ${\mathbf r}_i^{\hspace{0.05cm} \rm L}$ is the coordinate of the $i$th particle in the laboratory frame. The operators $A$ raise a particle two shells up, $B$ are the conjugate lowering operators, and $C$ are the generators of U(3), including the scalar $H^{(00)}$ that counts the total number of HO quanta. 
In the $z$-th direction, we have
\begin{equation}
    \frac{1}{b^2} \sum_i (z_i^{\hspace{0.05cm} \rm L})^2 = A_{zz} + B_{zz} + C_{zz}.
\end{equation}
Therefore,
\begin{subequations}
\begin{align*}
    \mathcal{I}_z^{\rm L} 
    &=  m \langle \sum_i [ (r_i^{\hspace{0.05cm} \rm L})^2 - (z_i^{\hspace{0.05cm} \rm L})^2 ] \rangle\\
    &= m b^2 \langle A_{xx} + B_{xx} + C_{xx} + A_{yy} + B_{yy} + C_{yy} \rangle.
\end{align*}
\end{subequations}
Using $C_{zz} = \hat N_z$ and $\sum_{\alpha} C_{\alpha \alpha} = \hat N$, where $\hat N$ ($\hat N_z$) is the operator of the total number of HO quanta (in the $z$ direction), we get
\begin{subequations}
\begin{align}
    \mathcal{I}_z^{\rm L} &= m b^2 \langle A_{xx} + A_{yy} + B_{xx} + B_{yy}\rangle 
                + m b^2 \langle \hat N - \hat N_z \rangle \notag\\
        &\simeq m b^2 \langle \hat N - \hat N_z \rangle,
\end{align}
\label{mozL}
\end{subequations}
in the laboratory frame. In the last step we use $\langle A_{xx} \rangle \simeq 0$, $\langle A_{yy} \rangle \simeq 0$, $\langle B_{xx} \rangle \simeq 0$, $ \langle B_{yy} \rangle \simeq 0$. These approximations follow from empirical observations; namely, \textit{ab initio} calculations show that excitations are favored first in the $z$ direction, and then in the $x$ direction, making $zz$ and $zx$ the dominant excitations (e.g., see \cite{launeydd16}). Almost no excitations occur in the $xx$ or $yy$ directions, allowing us to drop these terms.

\section{Removal of the Center-of-mass Contribution for $\mathcal{I}_z$}\label{appCM}
Because we use laboratory coordinates in our calculations, we must remove the spurious center-of-mass contribution for $\mathcal{I}_z$. That is, we need to calculate observables using intrinsic coordinates, ${\mathbf r}_i = {\mathbf r}_i^{\hspace{0.05cm} \rm L} -{\mathbf R}$, where  ${\mathbf R} = \frac{1}{A} \sum_i {\mathbf r}_i^{\hspace{0.05cm} \rm L}$ is the center-of-mass coordinate. Therefore, for the $z$-component of the moment of inertia of Eq. (\ref{momz}), we calculate 
\begin{align}
\Big\langle \sum_i {\mathbf r}_i^{\ 2}
- \sum_i z_i^{\ 2} \Big\rangle
&= \Big\langle \sum_i ({\mathbf r}_i^{\hspace{0.05cm} \rm L} - {\mathbf R})^2 - \sum_i (z_i^{\hspace{0.05cm} \rm L} - { R}_z)^2 \Big\rangle \notag \\
& \hspace{-1.75cm} = \Big\langle  \sum_i ({\mathbf r}^{\hspace{0.05cm} \rm L}_i)^2 - A {\mathbf R}^2  -  \sum_i (z^{\hspace{0.05cm} \rm L}_i)^2 + A { R}_z^2 \Big\rangle \notag\\
& \hspace{-1.75cm} = \Big\langle \sum_i \Big[ ({\mathbf r}^{\hspace{0.05cm} \rm L}_i)^2 - (z^{\hspace{0.05cm} \rm L}_i)^2 \Big] \Big\rangle -  A \Big\langle {\mathbf R}^2 - { R}_z^2 \Big\rangle.
\label{RelFromLab}
\end{align}
The first term is given by Eq. (\ref{mozL}). We now turn to calculating the second    term related to the center-of-mass (CM).

In the SA-NCSM eigenfunctions, the CM is exactly separated from intrinsic degrees of freedom: $\ket{\psi} = \ket{\phi^{\text{CM}}_{N=0,L=0,M=0}} \ket{\psi_{int}},$ where
\begin{equation}
   \phi^{\text{CM}}_{000} ({\mathbf R})=\langle {\mathbf R} \ket{\phi^{\text{CM}}_{000}} 
    = \frac{e^{ -R^2/2b_{\text{CM}}^2 }}{\pi^{1/4} b_{\text{CM}}^{3/2}} Y_{00}(\hat R)
\end{equation}
is the lowest HO center-of-mass wavefunction, and $b_{\text{CM}} = b/\sqrt{A}$. The CM operators in the second term of Eq. (\ref{RelFromLab}) only act on this CM wavefunction, so it suffices to calculate $\int |\phi^{\text{CM}}_{000} ({ R})|^2 R^4 dR$ and $\int |\phi^{\text{CM}}_{000} ({ R})|^2 R_z^2 R^2 dR$, which yield $3 b^2 / 2A $ and $b^2/2A$, respectively. Then,
\begin{align}
\Big\langle \sum_i \Big( {\mathbf r}_i^{\ 2} - z_i^{\ 2} \Big) \Big\rangle 
&= b^2 \langle \hat N - \hat N_z \rangle - \frac{3 b^2}{2} + \frac{b^2}{2} \notag\\
&= b^2 \langle \hat N - \hat N_z - 1\rangle,
\end{align}
and $\mathcal{I}_z$ is corrected by only a constant:
\begin{equation}
    \mathcal{I}_z = m b^2 \langle \hat N - \hat N_z - 1\rangle.
\end{equation}

Remarkably, $\mathcal{I}_z$ is diagonal in the SU(3) basis with  $\hat N \ket{N(\lambda \, \mu)}=N \ket{N(\lambda \, \mu)}$ and $\hat N_z \ket{N(\lambda \, \mu)}=N_z \ket{N(\lambda \, \mu)}$ with $N_z={1 \over 3} (2 \lambda + \mu + N)$. Hence, the translationally invariant $\mathcal{I}_z$ can be straightforwardly calculated for nuclear wavefunctions $\ket{J^\pi}$ in the SU(3) basis (\ref{SAbasis}) with probability amplitudes $(c^J_{N(\lambda \, \mu)})^2$:
\begin{equation}
    \mathcal{I}_z = m b^2 \sum_{N(\lambda \, \mu)} (c^J_{N(\lambda \, \mu)})^2 (N -  N_z - 1).
\end{equation}

}

\bibliography{main}

\end{document}